\documentclass[11pt]{revtex4}
\usepackage{amssymb,epsf}
\usepackage{latexsym}
\begin{document}

\title{Interaction Properties of the Periodic and Step-like Solutions of the
Double-Sine-Gordon Equation}
\author{M. Peyravi$^{1}$, Afshin Montakhab$^{2}$\footnote{email: montakhab@shirazu.ac.ir}, N. Riazi$^{1}$\footnote{email: riazi@physics.susc.ac.ir}
and A. Gharaati$^{3}$ } \affiliation{$1$. Physics Department and
Biruni Observatory
, Shiraz University, Shiraz 71454, Iran,\\
$2$. Physics Department, Shiraz University, Shiraz 71454, Iran,\\
$3$. Physics Department, Payame Noor University, Shiraz, Iran.}

\begin{abstract}
The periodic  and step-like solutions of the double-Sine-Gordon
equation are investigated, with different initial conditions and
for   various  values of the potential parameter $\epsilon$. We
plot energy and force diagrams, as  functions of the inter-soliton
distance for such solutions. This allows us to consider our system
as an interacting many-body system in $1+1$ dimension. We
therefore plot state diagrams (pressure vs. average density) for
step-like as well as periodic solutions. Step-like solutions are
shown to behave similarly to their counterparts in the Sine-Gordon
system. However, periodic solutions show a fundamentally different
behavior as the parameter $\epsilon$ is increased. We show that
two distinct phases of periodic solutions exist which exhibit
manifestly different behavior. Response functions for these phases
are shown to behave differently, joining at
an apparent phase transition point.     \\ \ \\

PACS: 05.45.Yv, 05.00.00, 02.60.Lj, 24.10.Jv
\end{abstract}

\maketitle

\section{Introduction\label{Intr}}

The Sine-Gordon (SG) equation is a non-linear partial differential
equation which appears naturally in different physical systems in
atomic physics \cite{0}, electromagnetism\cite{1},
superconductivity\cite{2}, field theory\cite{3},
biophysics\cite{4,5,6}, and statistical mechanics\cite{7}.

  The double-Sine-Gordon (DSG) equation which is a
generalization of the ordinary SG equation has been the focus of
much recent investigations. It has been shown to model a variety
of systems in condensed matter, quantum optics, and particle
physics\cite{8}. Condensed-matter applications include the spin
dynamics of superfluid $^{3}He$\cite{9,10}, magnetic
chains\cite{11}, commensurate-incommensurate phase
transitions\cite{12}, surface structural reconstructions\cite{13},
domain walls\cite{14,15} and  fluxon dynamics in Josephson
junction\cite{15b}.

\begin{figure}
\epsfxsize=10cm\centerline{\epsfbox{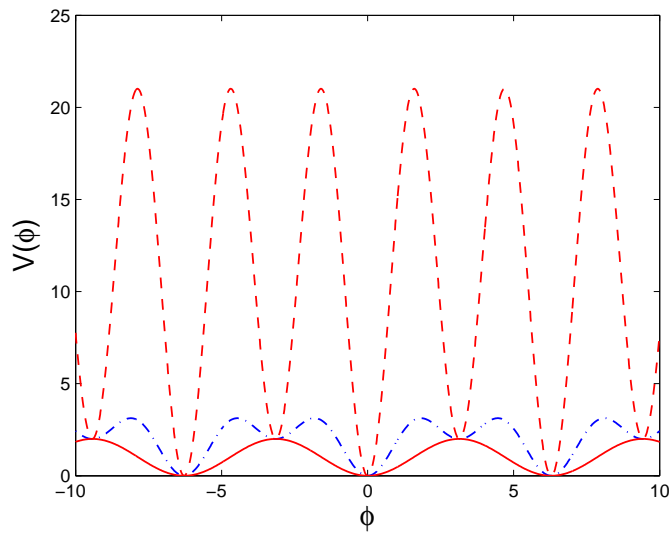}} \caption{DSG
Potential. The dashed curve is for $\epsilon=10$, the dash-dotted
curve is for $\epsilon=1$ and the solid curve is for
$\epsilon=0$(SG).\label{s}}
\end{figure}

 In quantum field theory and quantum optics DSG
applications include quark confinement\cite{16} and self-induced
transparency\cite{17}. The internal dynamics  of multiple and
single DSG soliton configurations using molecular dynamics have
been studied in\cite{8}. There have also been studies about kink
anti-kink collision processes for DSG equation \cite{18}. One can
also point to the statistical mechanical applications\cite{19},
and perturbation theory for this equation\cite{20}. It should be
mentioned that the potentials adopted in various studies are not
exactly the same. The DSG potential which contains a constant and
a harmonic term in addition to the self-interaction potential of
the ordinary SG equation is considered here\cite{20,21}:
\begin{equation}
V(\phi) = 1 + \epsilon - \cos\phi -\epsilon\cos(2\phi).
\end{equation}
where $\epsilon$ is a constant. This potential is sketched in
Fig.\ref{s} for $\epsilon=0, 1$ and $10$. Some dynamical
properties of multiple and single soliton
 solutions of the DSG system were studied by Burdick et. al\cite{8},
 using molecular dynamics.

 Following the common terminology
in field theory, this potential has absolute degenerate minima at
$\phi = 2n\pi$ as the true vacuua, and the metastable, local
minima at $\phi = (2n+1)\pi$ as the false vacuua\cite{21}. False
vacuua  develop if $\epsilon > 0.25$\cite{21}. The harmonic term
in this potential can result from the Fourier expansion of an
arbitrary, periodic potential $\widetilde{V}(\phi) =
\widetilde{V}(\phi+2n\pi)$. One does not expect the system to
remain integrable by adding these extra terms\cite{21}. The
potential reduces to the ordinary SG potential in the limit
$\epsilon\longrightarrow0$.

In this article, we consider chains of the kink and anti-kink
solutions of DSG equation in $1+1$ dimension. Our goal is to
consider a one dimensional chain of many solitons and treat them
as a many-body interacting system. As will be shown shortly, these
solutions do have mutual interactions which is a function of their
mutual distance. We thus propose to study the DSG system as a
many-body  interacting system at zero temperature in one spatial
dimension.

The structure of our presentation is as follows: in Section
\ref{Field} we review some basic properties of the DSG system. In
Section \ref{Field1}, we study the periodic and step-like
solutions of the DSG system, highlighting their interaction
properties. In Section \ref{Field2}, we study the macroscopic
properties of such systems viewed as a many-body interacting
system. We close in Section \ref{Field3} by summarizing our
results and pointing some directions for future work.

\section{Basic Properties of the Double Sine-Gordon System} \label{Field}

From a relativistic point of view, the double-Sine-Gordon
Lagrangian density can be put in the following form:
\begin{equation}\label{a}
{\cal L}_{DSG}= \frac{1}{2}\partial^{\mu}\phi\partial_{\mu}\phi-[1
+ \epsilon-\cos\phi-\epsilon\cos(2\phi)].
\end{equation}
From this Lagrangian density, we obtain the following equation:
\begin{equation}
\Box\phi = -\sin\phi - 2\epsilon\sin(2\phi);
\end{equation}
for the real scalar field $\phi(x,t)$ in $(1+1)$ dimensions. It
can be shown that this equation has the following exact, static,
single kink (anti-kink) solution\cite{21}:
\begin{figure}
\epsfxsize=9cm\centerline{\epsfbox{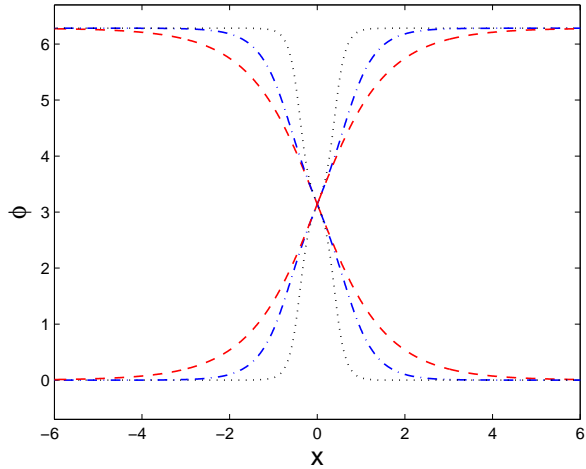}}\caption{Kink and
anti-kink solutions of DSG. The dotted curves are for
$\epsilon=10$, the dash-dotted curves are for $\epsilon=1$ and the
dashed curves are for $\epsilon=0$(SG).\label{h}}
\end{figure}

\begin{eqnarray}  \label{b}
\phi(x) = 2\arccos\left[\pm
\frac{\sinh\sqrt{4\epsilon+1}x}{\sqrt{4\epsilon+\cosh^{2}\sqrt{4\epsilon+1}x}}\right].
\end{eqnarray}

\begin{figure}
\epsfxsize=9cm\centerline{\epsfbox{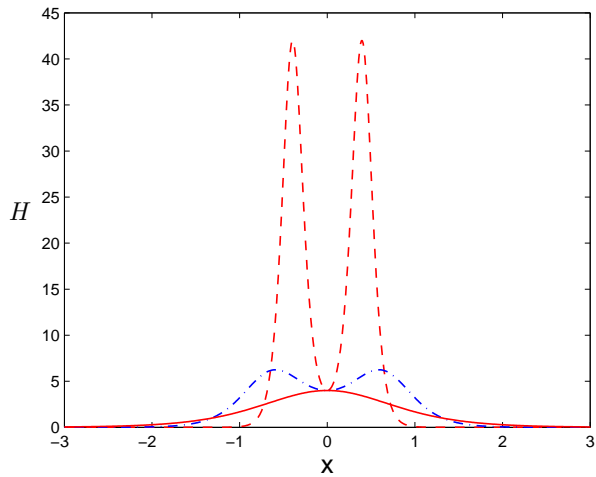}}\caption{DSG  energy
density for the kink solution. The dashed curve is for
$\epsilon=10$, the dash-dotted curve is for $\epsilon=1$ and the
solid curve is for $\epsilon=0$(SG). It can be seen that sub-kinks
develop for large enough values of $\epsilon$ ($\epsilon>0.25$)
\label{m}}
\end{figure}

 These are plotted in Fig.\ref{h} for three values of the parameter $\epsilon(=0, 1,
 10)$. Note that the $(+)$ sign is for the anti-kink ($\tilde{k}$), while the $(-)$
sign is for the kink ($k$). The solution (\ref{b}) tends to the
conventional kink (anti-kink) solution  of the SG equation in the
limit $\epsilon\longrightarrow0$. As the value of $\epsilon$
grows, the kink develops two parts (or sub-kinks)\cite{21}, see
Fig.\ref{m}.

Exact solutions of DSG system with arbitrary constant coefficients
were obtained recently by Wang and Lie\cite{21b} by F-expansion
method which is a generalization of the Jacobi elliptic function
expansion. It is obvious that solitary and periodic solutions of
the SG system are special cases of these solutions (with
particular choices of the constants of integration). The
thermodynamic properties of the DSG chain were studied in Condat
et. al\cite{21c}, in which the polarization precursor was
determined.

 Using the Noether's theorem and the invariance of the action under the the space-time translation
 $x^{\mu}\rightarrow^{\mu}+a^{\mu}$, one easily obtains\cite{22,23}:
\begin{equation}
T^{\mu\nu} = \partial^{\mu}\phi\partial^{\nu}\phi -
g^{\mu\nu}{\cal L}_{DSG};
\end{equation}
in which $g^{\mu\nu}=diag(1,-1)$ is the metric of the $(1+1)$
dimensional Minkowski spacetime. The energy-momentum tensor of the
true vacuum $(\phi=2n\pi)$ vanishes, while for
 the false vacuum $(\phi=(2n+1)\pi)$, it is non-vanishing\cite{21}:
 \begin{equation}\label{qq}
 T^{\mu\nu}=2g^{\mu\nu}.
 \end{equation}
 As expected, this tensor is that of a perfect fluid, with the
 equation of state $p=-\rho=-2$\cite{21}, similar to that of the
 cosmological constant\cite{23}. Using the virial theorem,
 the energy density of the static kink is given by (see Fig.\ref{m}):
 \begin{equation}
{\cal
H}(x)=2V(\phi)=2[1+\epsilon-\cos(\phi(x))-\epsilon\cos(2\phi(x))].
\end{equation}
As for the SG system, the topological current for the DSG system
is given by\cite{24}:
\begin{equation}\label{c}
J^{\mu}=\frac{1}{2\pi}\epsilon^{\mu\nu}\partial_{\nu}\phi;
\end{equation}
 where $\epsilon^{\mu\nu}$ is the totally antisymmetric tensor in
 two dimensions. The current density (\ref{c}) is conserved:
 \begin{equation}
 \partial_{\mu}J^{\mu}=\frac{1}{2\pi}\epsilon^{\mu\nu}\partial_{\mu}\partial_{\nu}\phi=0.
 \end{equation}
 since $\epsilon^{\mu\nu}$ is antisymmetric while
 $\partial_{\mu}\partial_{\nu}$ is symmetric in the indices.
  The total topological charge of a localized solution is given by:
 \begin{eqnarray}
 Q=\int_{-\infty}^{+\infty}J^{0}dx =  \frac{1}{2\pi}\int_{-\infty}^{+\infty}\epsilon^{01}\partial_{1}\phi
 dx \nonumber \\= \frac{1}{2\pi}\int_{-\infty}^{+\infty}\frac{\partial\phi}{\partial
 x} dx = \frac{1}{2\pi}[\phi(+\infty)-\phi(-\infty)].
 \end{eqnarray}
 Considering the boundary conditions of a localized solution, the DSG equation leads to:
 \begin{equation}
 \phi(+\infty)=2n\pi ,  \qquad  \phi(-\infty)=2m\pi;  \qquad
 Q=n-m.
 \end{equation}
Localized, static solutions have $n-m=\pm2\pi$, and therefore
  \begin{equation}\label{6}
  Q=\pm1      \qquad  \text{for kink (anti-kink)}.
  \end{equation}
  As the value of $\epsilon$ grows, sub-kinks form, which have
  half-integral topological charges\cite{21}.

  The stability of kink solutions of the DSG system is guaranteed
  by topological reasons. The kink is the lowest energy
  configuration with the prescribed  boundary conditions and the
  topological charge given by Eq.(\ref{6}).

\section{Periodic and Step-Like solutions}\label{Field1}

 In this section, we classify and study two types of static
 solutions of the DSG equation, for which we have:

\begin{equation}\label{1}
\frac{d^{2}\phi}{dx^{2}}=\frac{dV(\phi)}{d\phi}.
\end{equation}
In order to proceed, we use a Runge-Kutta method\cite{24a} to
integrate Eq. (\ref{1}) numerically. We choose $\phi=\pi$ and
various values of $\frac{d\phi}{dx}$ as our initial conditions at
$x=0$. We study this system for various potential parameters
$\epsilon$. We choose $\epsilon=1$ as a typical value but also
compare results with $\epsilon=0.1$ (SG-like behavior) as well as
$\epsilon=10$ for decidedly different behavior. It is easy to see
that solutions are characterized by an (x-independent) constant
$P$:
\begin{equation}\label{2}
P=\frac{1}{2}\left(\frac{d\phi}{dx}\right)^2-V(\phi);
\end{equation}
\begin{equation}
\frac{dP}{dx}=0.
\end{equation}
Note that this is different from energy density
\begin{equation}\label{8}
{\cal H}(x)=\frac{1}{2}\left(\frac{d\phi}{dx}\right)^2+V(\phi);
\end{equation}
which changes with position $x$. We will interpret $P$ as
``pressure'' (or tension in this 1d case) for reasons to be
explained in the next section.

\begin{figure}[h]
\epsfxsize=10cm\centerline{\epsfbox{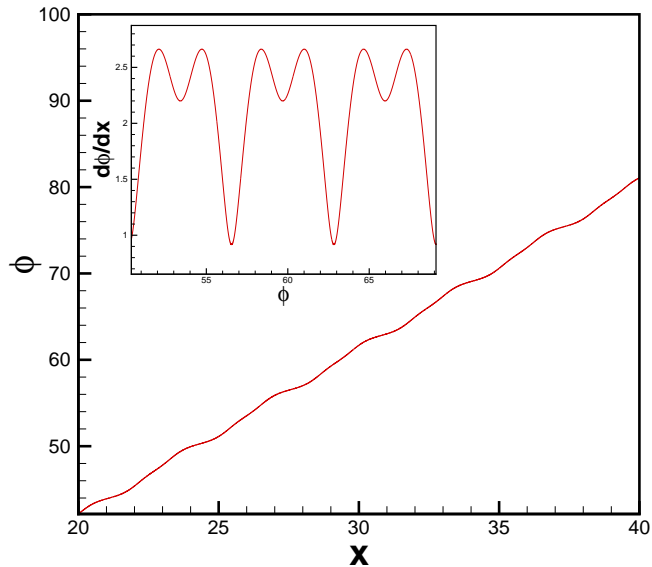}}\caption{Step-like
chain of DSG solitons for $\epsilon=1$  and $P=0.42$. The inset
shows the cyclic nature of these solutions in more
details.\label{go}}
\end{figure}
\begin{figure}[h]
\epsfxsize=10cm\centerline{\epsfbox{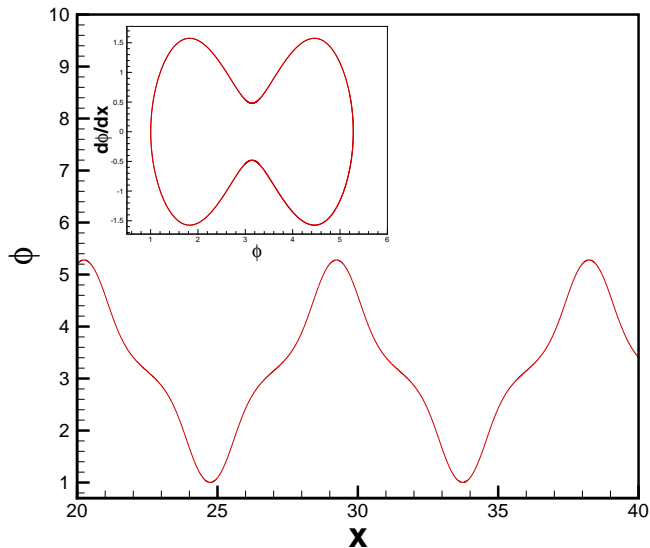}}\caption{ Periodic
chain of DSG solitons for $\epsilon=1$ and $P=-1.8848$. The inset
shows the cyclic nature of these solutions in more
details.\label{co}}
\end{figure}
Two types of solutions emerge under such conditions:  Step-like
solutions and periodic solutions (see Fig.\ref{go} and \ref{co}).
Step-like solutions are characterized by $P>0$ and periodic
solutions are characterized for $-2<P<0$. Note that this
classification is independent of $\epsilon$ and therefore holds
true also for SG equation\cite{24b}. As can be seen from the
relevant figures the step-like solutions are a sequence of kinks
($kkk...$) or anti-kinks ($\tilde{k}\tilde{k}\tilde{k}$...), and
the periodic solutions are a sequence of kink, anti-kink
($k\tilde{k}k\tilde{k}...$) solutions. A simple phase diagram of
Eq.~(13) is shown in Fig.~6.

\begin{figure}[h]
\epsfxsize=10cm\centerline{\epsfbox{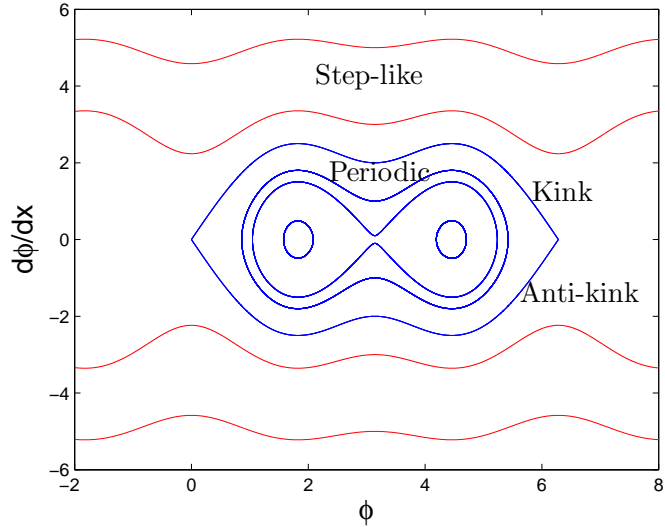}}\caption{A typical
phase diagram for solutions of Eq.~(13) with  $\epsilon=1.0$. The
two different solutions are clearly marked on the diagram.\label{pd}}
\end{figure}

\begin{figure}[h]
\epsfxsize=9cm\centerline{\hspace{8cm}\epsfbox{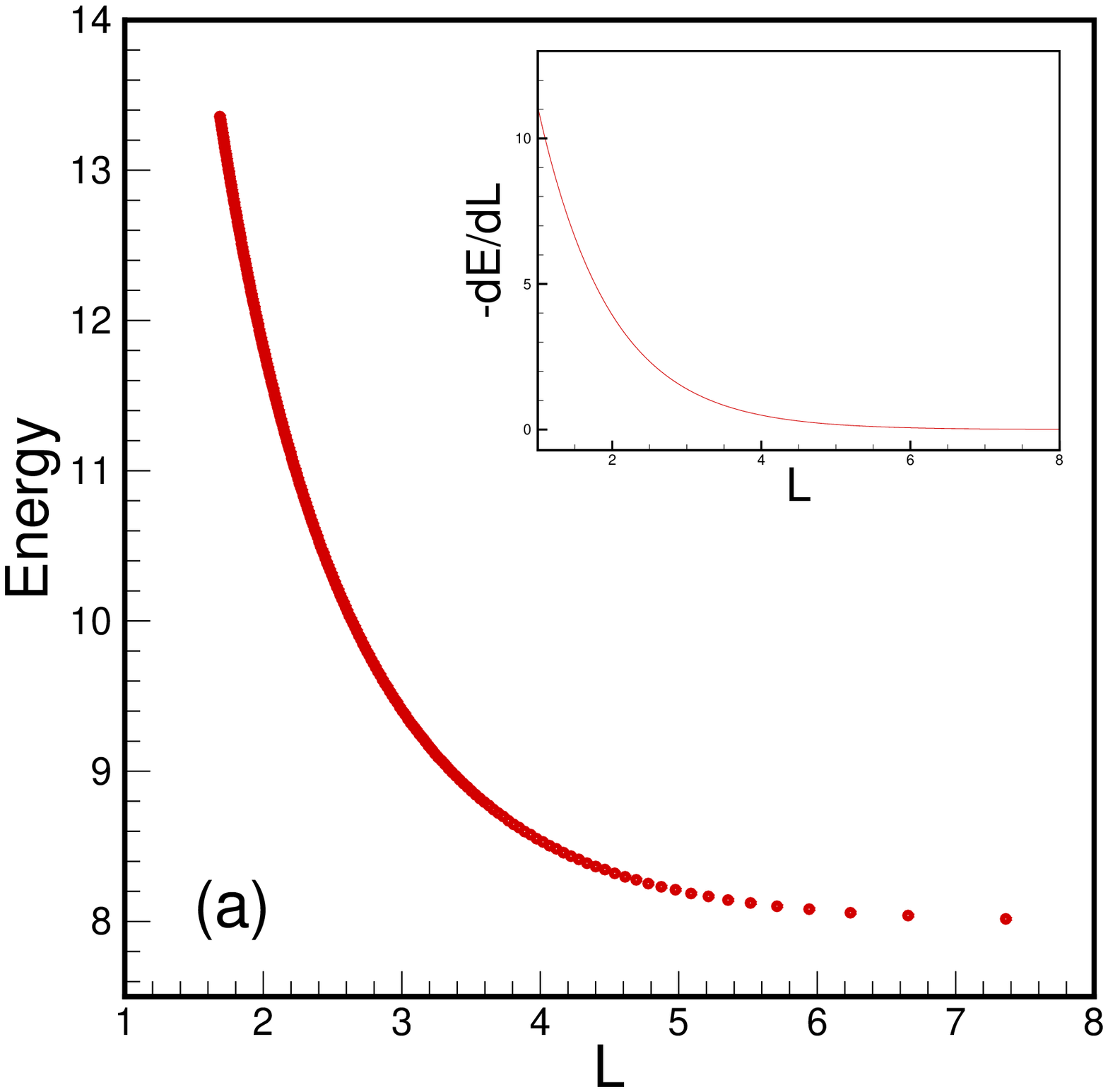}\epsfxsize=9cm\centerline{\hspace{-7cm}\epsfbox{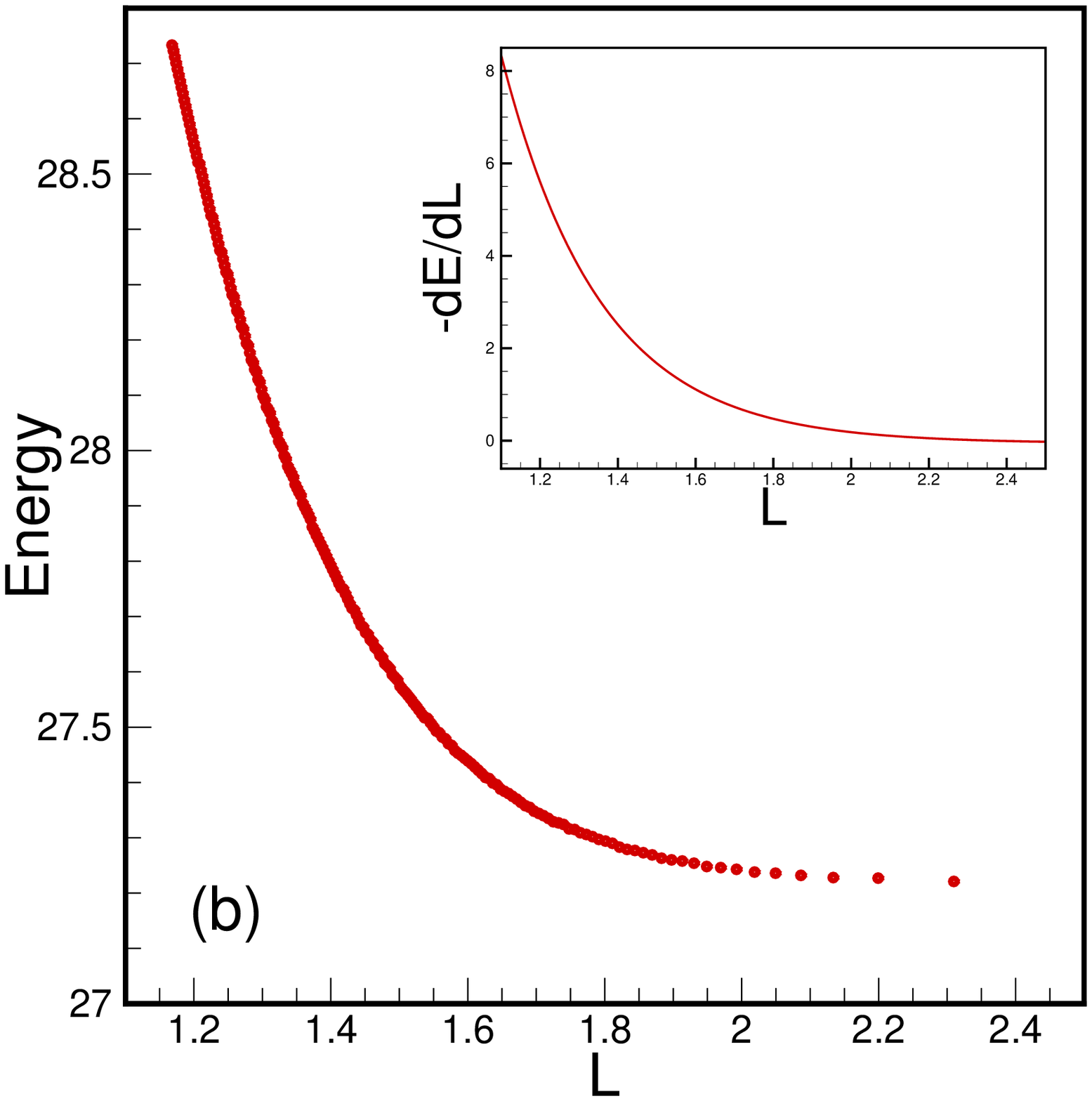}}
}\caption{Energy per soliton diagram for step-like chain of (a) SG
solitons and (b) DSG solitons for $\epsilon=10$. The behavior of
the system dose not change significantly as $\epsilon$ is changes.
The inset shows the derivative which we interpret as the
(repulsive) force between the solitons.\label{r}}
\end{figure}

Using Eq.(\ref{8}), one can characterize these solutions via an
energy function. We take a given static solution and compute the
total energy per ``soliton'' i.e. integral of Eq.(\ref{8}) over
one period. For each static solution we also calculate the
``inter-soliton'' distance, L, as a distance between two
successive (similar) peaks of energy density. We can therefore
plot energy per soliton as a function of inter-soliton distance
for various values of $\epsilon$ and P. Since it is our goal to
consider these systems as a many-body system, we use energy as a
function of distance E(L) and compute the ``inter-soliton force''
$F=-\frac{dE}{dL}$. Figure \ref{r} shows such results for the
step-like solutions. No significant change of behavior is seen as
$\epsilon$ is varied for these type of solutions. We note that in
both cases (SG and DSG) at short distances the energy rises
abruptly reminiscent of hard core potentials. It quickly falls as
distance $L$ is increased reaching an asymptotic value for large
$L$. Accordingly, the ``inter-soliton force'' $F$ (insets) is
repulsive and falls to zero quickly as distance is increased,
indicating a short-range interaction.

\begin{figure}[h]
\epsfxsize=9cm\centerline{\epsfbox{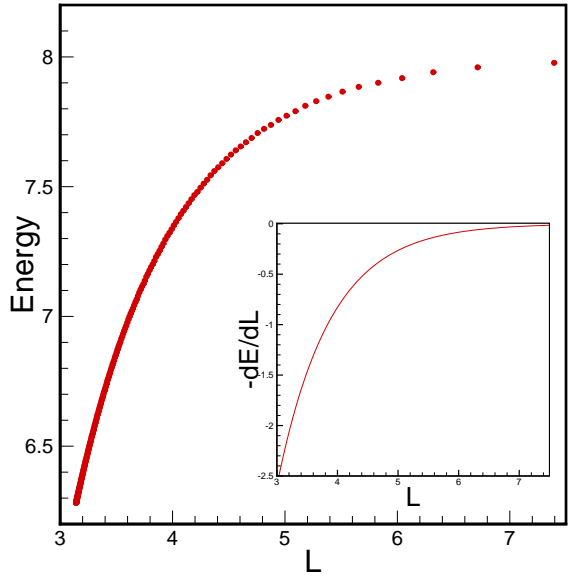}}\caption{Energy per
soliton diagram for periodic chain of the  SG solitons. The inset
shows the derivative which we interpret as the (attractive) force
between the solitons.\label{p}}
\end{figure}
\begin{figure}[h]
\epsfxsize=18cm\centerline{\epsfbox{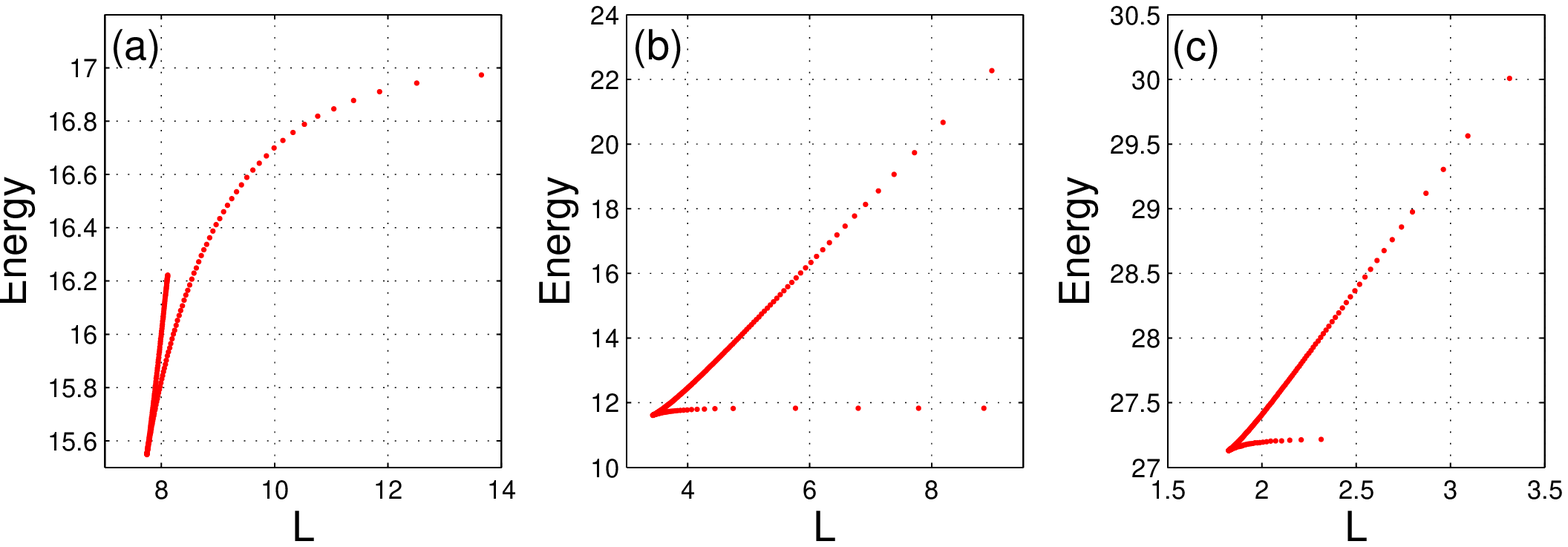}}\caption{Energy per
soliton diagram for periodic chain of DSG solitons (a) for
$\epsilon=0.1$, (b) for $\epsilon=1$ and (c) for $\epsilon=10$.
Note the emergence of a new (upper) branch as $\epsilon$ is
increased. The lower branch reaches a constant value for large
$L$, but the upper branch increases linearly with slope
2.\label{e}}
\end{figure}

\begin{figure}[h]
\epsfxsize=9cm\centerline{\epsfbox{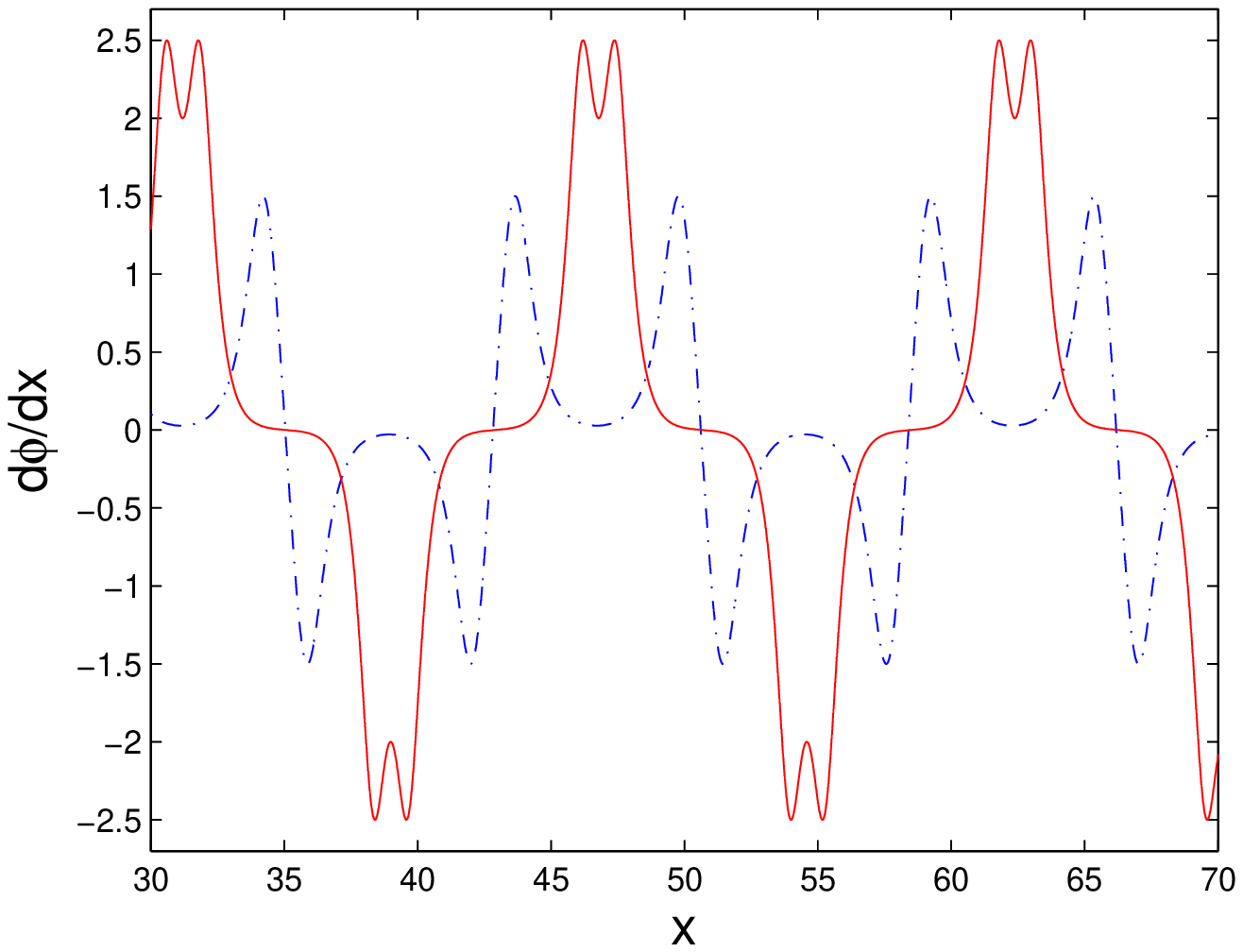}}\caption{The slope
diagram for the periodic chain of DSG  solitons for $\epsilon=1$ ,
with $P= -2.14\times10^{-5} $ for the curve with larger amplitude
(solid curve) and with $P=-1.9996$ for the curve with smaller
amplitude (dash-dotted curve). The two solutions have the same
inter-soliton distance (L=7.798) but different energy per
soliton.\label{AB}}
\end{figure}

\begin{figure}[h]
\epsfxsize=9cm\centerline{\hspace{8cm}\epsfbox{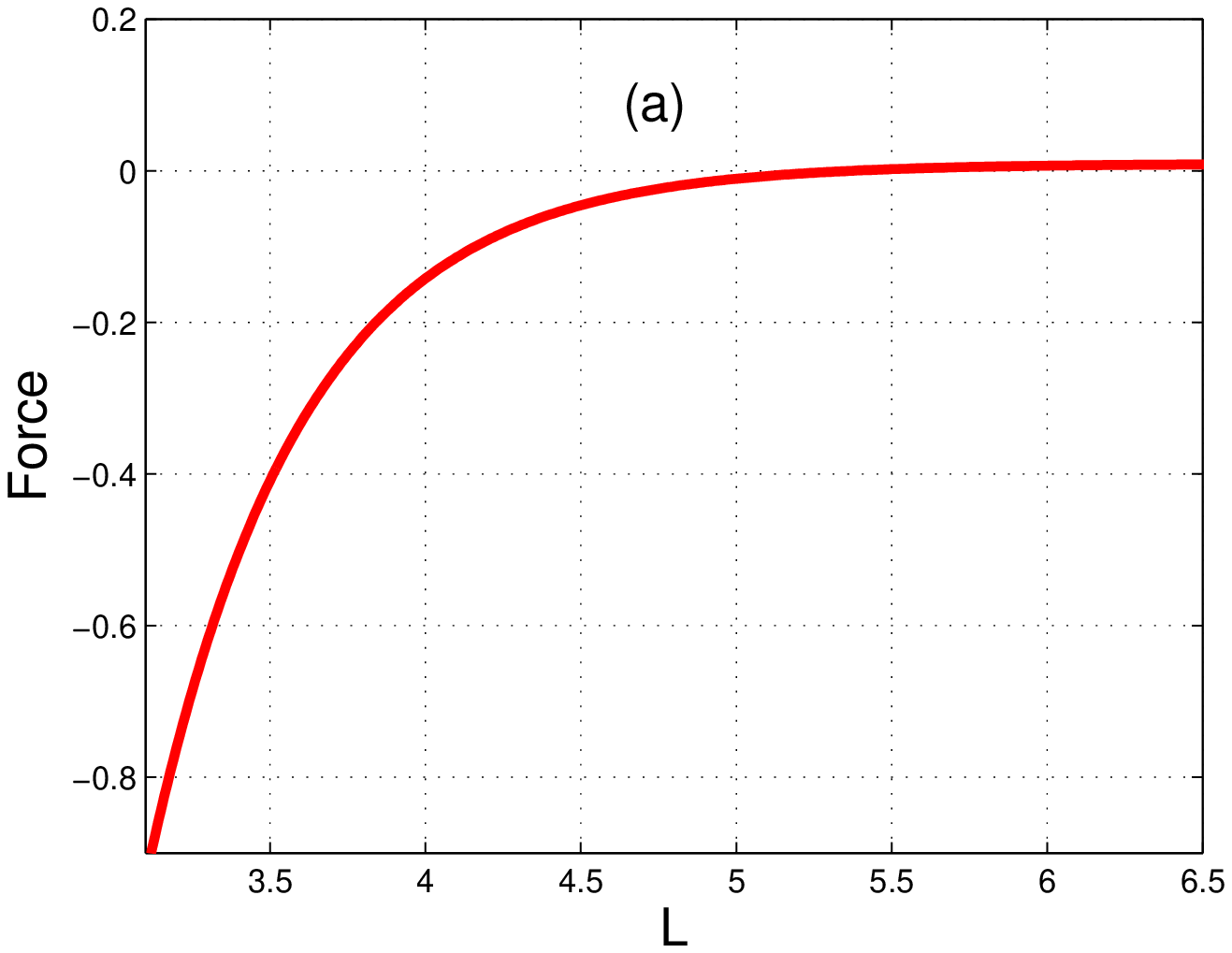}\epsfxsize=9cm\centerline{\hspace{-7cm}\epsfbox{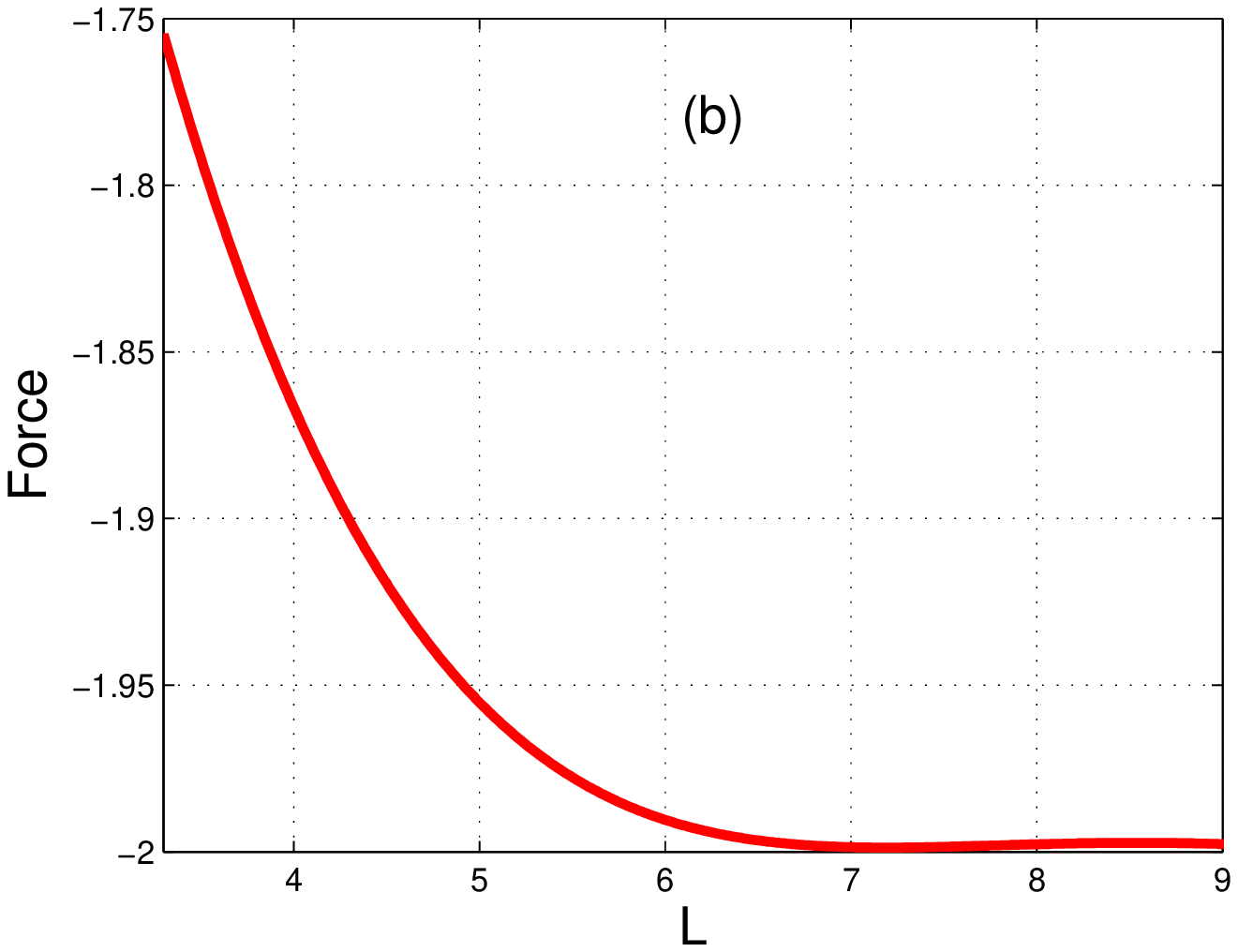}}}
\caption{Attractive force for: (a) the lower branch and (b) the
upper branch of periodic chain of DSG solitons for
$\epsilon=1$.\label{shp}}
\end{figure}

We next consider the periodic solutions. Here, as we will show,
$\epsilon$ changes the behavior of these solutions in a
significant way. Figure \ref{p} shows the results for the periodic
solutions for $\epsilon=0$ (SG). The behavior is the reverse of
the step-like solutions. The energy decreases sharply at short
distances and increases quickly to a constant value at large $L$.
Therefore, the inter-soliton force is attractive and falls off to
zero at large distances (inset). What happens as $\epsilon$ is
changed in this case? The results for the energy diagram is shown
in Fig.\ref{e}. As is seen from the figure, the introduction of
$\epsilon$ leads to the emergence and growth of a new branch of
solutions. It is important to distinguish between these two types
of periodic solutions. The inter-soliton distance $L$ changes with
$P$ which is a function of initial conditions. In the lower branch
(the SG-like branch), as $L$ becomes large, the energy per soliton
tends to its single-soliton  value, thus leading to an independent
soliton system. However, in the upper branch, as $L$ become large
the sub-kinks also separate out with a false vacuum energy
developing between them. Here, as $L$ grows large, the total
energy in the false vaccua increases linearly in the space between
the sub-kinks, which is indicated by the linear dependence of
energy in Fig.\ref{e}\cite{25a}. To help visualize these different
solutions, we show in Fig.\ref{AB} a periodic solution from each
branch, having the same $L$ but different $E$. The inter-soliton
force for the periodic solution is still attractive since in
either branch the energy is an increasing function of distance.
However, in the lower (SG like) branch the force falls off to zero
at large distances leading to a non-interacting soliton picture.
Note that this is the case in all the solutions considered thus
far. But, in the new upper branch, the attractive force increases
its (absolute) values to $-2$ as $L$ is increased (Fig.10b). This
is due to the above-mentioned linear behavior of energy in the
upper branch, caused by separation of sub-kinks with increasing
$L$. A typical force diagram for periodic solution of DSG system
with $\epsilon=1$ is shown in Fig.\ref{shp}

\section{Periodic and step-like solutions as a many-body system}\label{Field2}

In order to gain a better understanding of the soliton
interpretation and mutual interactions between solitons, one may
also study state diagrams which depict pressure $P$  versus
average density $\bar{\rho}$, where $P$ and $\rho$ are given
by\cite{25}:
\begin{equation}
P=-T^{1}_{1}=\frac{1}{2}\left(\frac{\partial\phi}{\partial
x}\right)^{2}-V(\phi).
\end{equation}

\begin{equation}
\rho=T^{0}_{0}=\frac{1}{2}\left(\frac{\partial\phi}{\partial
x}\right)^{2}+V(\phi).
\end{equation}

It should be noted that the energy-momentum tensor of the system
(like many other relativistic continuous media) takes the form
\begin{equation}
(T^{\mu}_{\nu})=\left(%
\begin{array}{cc}
  T^{0}_{0} & 0 \\
  0 & T^{1}_{1} \\
\end{array}%
\right)
\end{equation}
reminiscent of the energy-momentum tensor of a perfect fluid in
$1+1$ dimensions given by\cite{25}:

\begin{equation}
(T^{\mu}_{\nu})=\left(%
\begin{array}{cc}
  \rho c^{2} & 0 \\
  0 & -P \\
\end{array}%
\right)
\end{equation}
Note that pressure is the same as our first integral $P$
(Eq.(\ref{2})) and density $\rho$ is in fact the same as energy
(Hamiltonian) density (Eq.(\ref{8})). Here we use the generic word
``pressure'' as a thermodynamic conjugate variable to the size of
the system. If the system is three dimensional, pressure has its
usual meaning. However, for one dimensional systems like the ones
we are considering here ``tension'' is a more appropriate name for
our variable, ($\tau dl$ instead of $-Pdv$ as mechanical energy or
work). With this in mind, we use the word pressure or tension
interchangably.

\begin{figure}[h]
\epsfxsize=9cm\centerline{\hspace{8cm}\epsfbox{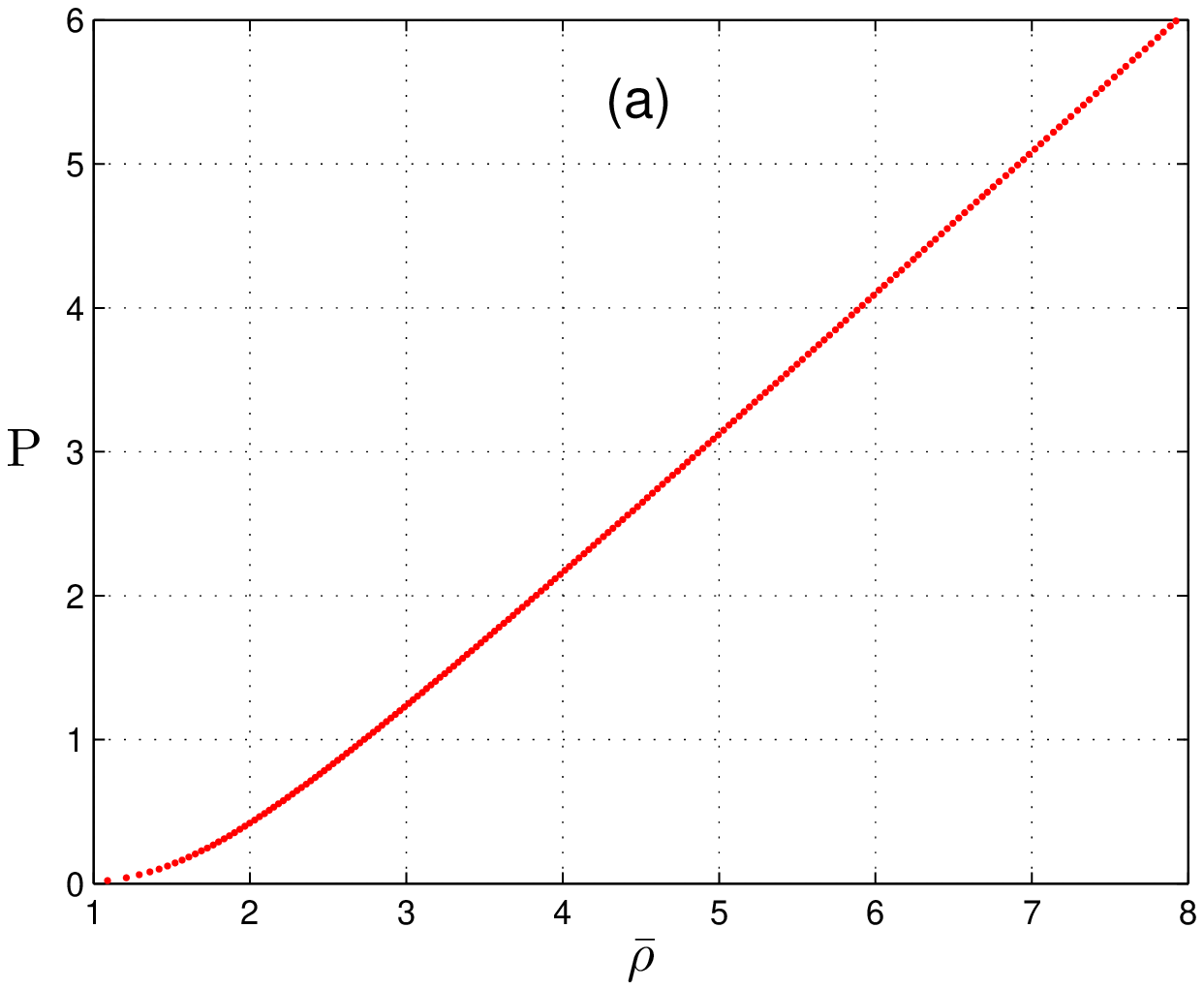}\epsfxsize=9cm\centerline{\hspace{-7cm}\epsfbox{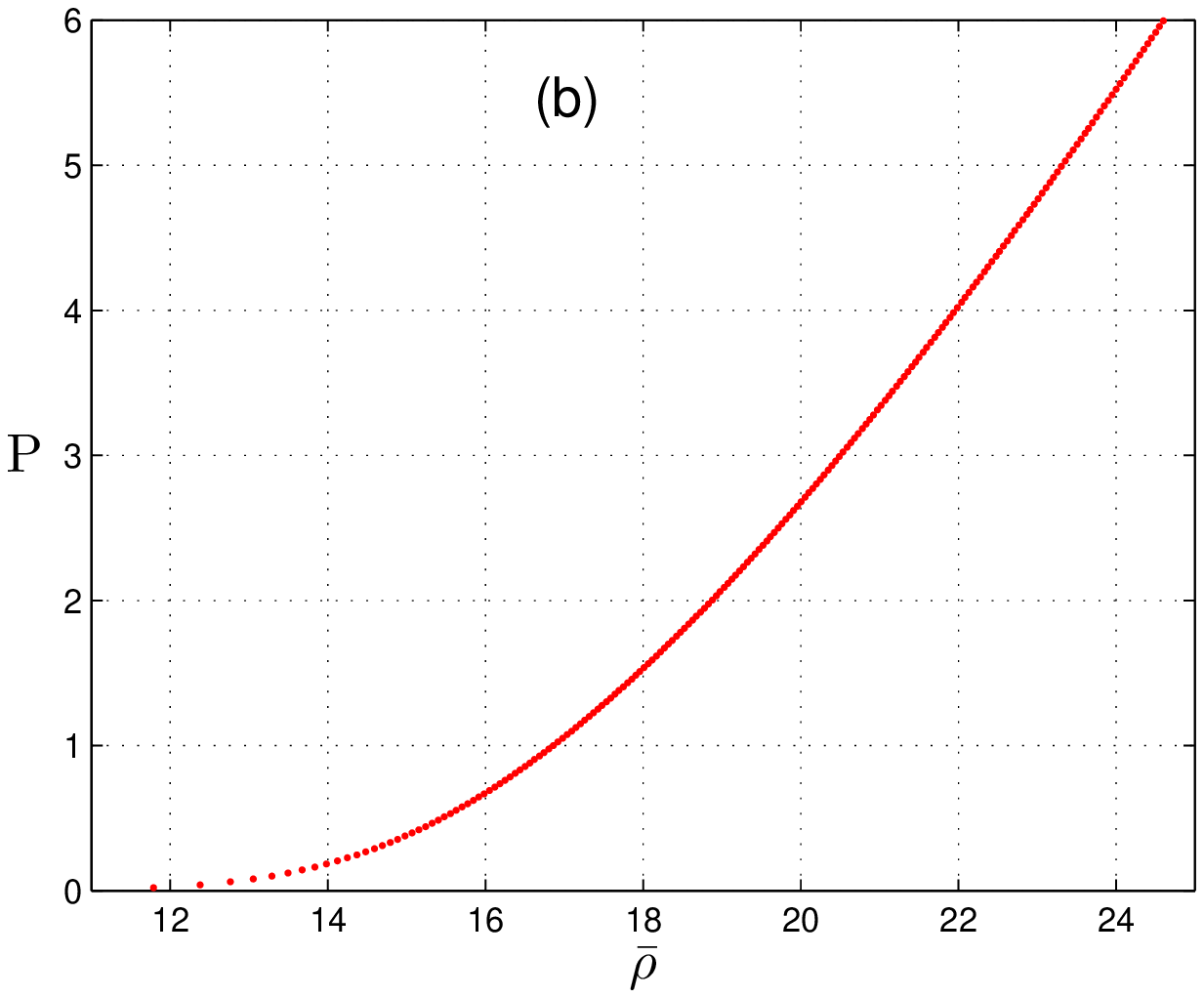}}
}\caption{Equation of state diagram for step-like chain of (a) SG
solitons and (b) DSG solitons for $\epsilon=10$. \label{no}}
\end{figure}

 In Fig.\ref{no}, we show a state diagram for SG as well as
DSG for $\epsilon=10$ for the step-like solutions. We note that,
as before, the behavior is essentially the same. The pressure
increases smoothly from zero and tends to a linear regime with
increasing density. This behavior is consistent with the repulsive
force diagram (see Fig.\ref{r}) which shows that the inter-soliton
force becomes stronger at short distances which indicates that
pressure increases with increasing density.
 \begin{figure}[h]
\epsfxsize=9cm\centerline{\epsfbox{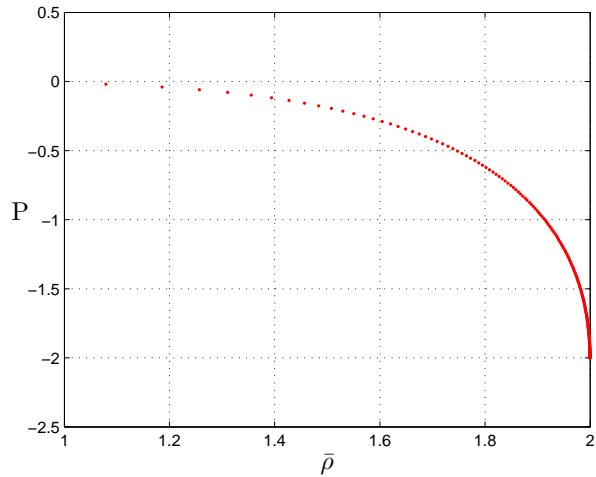}}\caption{Equation of
state diagram for periodic chain of SG solitons. \label{to}}
\end{figure}

\begin{figure}[h]
\epsfxsize=9cm\centerline{\epsfbox{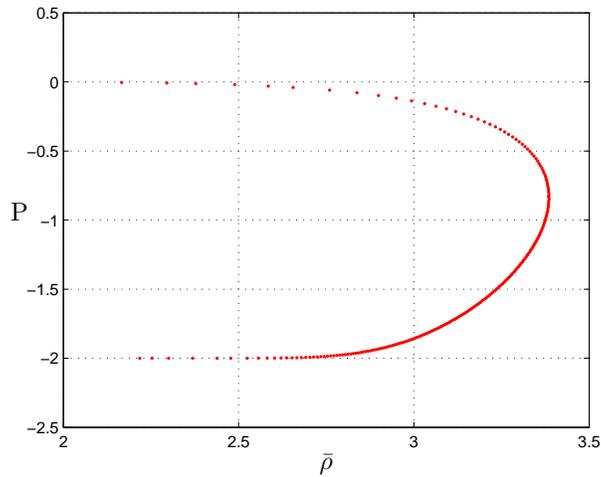}}\caption{Equation of
state diagram for periodic chain of DSG solitons for
$\epsilon=1$. For examples of solutions from each branch
see Fig.~10.\label{mo}}
\end{figure}

As the energy and force diagrams indicate, the state diagrams for
periodic chains are more complicated (and therefore more
interesting) than the step-like solutions. In Fig.\ref{to}, we
show the state diagram for a periodic chain of SG solitons. Here,
with increasing density, the pressure \textit{decreases} from
zero. Clearly this negative pressure is a result of attractive
force (see Fig.\ref{p}) among periodic kink-antikink SG solitons.
Note that as density increases and inter-soliton distance
decreases, the attractive force becomes stronger thus lowering the
pressure or increasing the \textit{tension} in this one
dimensional chain. The negative value of pressure is simply
related to the negative inter-soliton force. An interesting
feature of this diagram occurs at $\bar{\rho}\approx2$ which is
the maximum density. Here, $P=-2$ and $\frac{\partial
\bar{\rho}}{\partial P}=0$, i.e. the ``fluid'' becomes
incompressible. This simply corresponds to an exceedingly large
(attractive) force at short distances. We also note that since
$\frac{\partial \bar{\rho}}{\partial P}$ is negative, the term
``inexpandable fluid'' is perhaps more appropriate than
incompressible fluid. Note that the end point $(P,\rho)=(-2,2)$
exactly corresponds to the vacuum equation of state $P=-\rho$,
coming from Eq.(\ref{qq}). Note, however, that this point is
highly unstable in the SG system, since it corresponds to a local
maximum in Fig.\ref{s}.

Next, we study the state diagrams for a periodic DSG system. As
shown above, the energy diagram splits into two (upper and lower)
branches (see Fig.\ref{e}). Figure \ref{mo} shows a typical state
diagram for DSG periodic chain. In fact, the state diagram also
contains two (connected) branches. The upper half which is similar
to the previous state diagram for the simple SG system, as well as
a new, lower half which returns the same values of density but at
lower pressure values or larger tension. We note that the upper
half corresponds to the lower (SG-like) energy branch, whereas the
lower half corresponds to the new upper energy branch of DSG
system. Again, it is easy to understand this state diagram in
terms of its corresponding force diagram. Here, in the new lower
branch, increasing the density (and thus reducing $L$) causes a
weaker attractive force (see Fig.\ref{shp}) which in turn causes
an increase in the pressure.

\begin{figure}[h]
\epsfxsize=9cm\centerline{\epsfbox{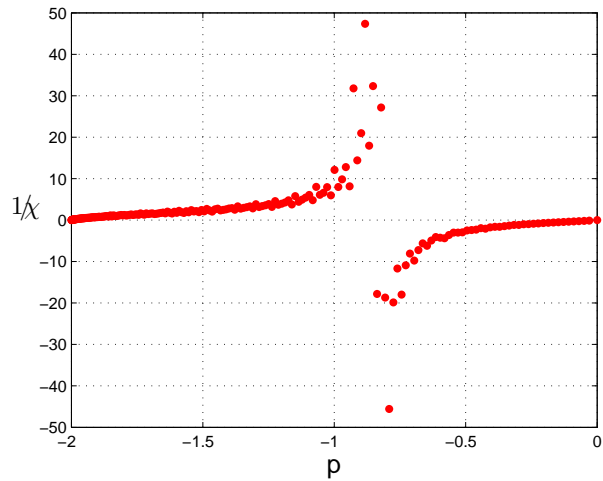}}\caption{$\frac{1}{\chi}$
versus $P$, showing clearly two phases with positive and negative
values of $\frac{1}{\chi}$. \label{cv}}
\end{figure}

The shape of the state diagram, Fig.\ref{mo}, with
double-valuedness of pressure is reminiscent of Van der Waals
fluid. There, the multiple-valuedness of density as a function of
pressure with a corresponding change in the sign of
compressibility $\chi=\frac{\partial \bar{\rho}}{\partial P}$
signals the onset of a liquid-gas transition as temperature is
lowered\cite{26}. Our system is obviously very different from that
described by the Van der Waals equation of state. However, our
system exhibits two distinct solutions, one with a positive and
the other with negative $\chi$, being joined at a maximum density
$\bar{\rho}_{max}=3.385$. This point could be thought of as a
phase transition point. The transition point is the point at which
the ``fluid'' becomes ``inexpandable'', which separates regimes of
different signs of compressibility. In other words, the transition
point is the point at which a stable solution ($\chi>0$) gives way
to an (thermodynamically) unstable solution ($\chi<0$). This can
be better seen in Fig.\ref{cv} where we plot $\frac{1}{\chi}$ vs.
$P$ in order to show these two distinct phases.

Energy per soliton (Fig.\ref{e}), and equation of state diagrams
(Fig.\ref{mo}) for the periodic chain clearly show two branches.
These two branches correspond to two different phases of the
chain. The two phases differ in the following respect: The upper
branch of Fig.\ref{mo} corresponds to a chain
$...\tilde{k}k\tilde{k}k\tilde{k}...$ with solitons separated by a
region of true vacuum (zero energy density). This is why at large
inter-soliton distances the force (and thus tension) vanishes for
this branch. The lower branch, on the other hand, corresponds to
pairs of sub-kinks (with vanishing topological charge) separated
by regions of false vacuum (non-vanishing energy density). This
explains the asymptotic value $-2$ at low densities (large
inter-soliton distances, see fig.\ref{shp}(b)). It is interesting
to note that the energy stored in the false vacuum increases
linearly with the inter-soliton distance (at large distances)
which leads to a constant force (and thus tension). This
interesting behavior is reminiscent of the confinement phenomenon
in hadronic physics\cite{23}.

\section{Summary and Conclusions}\label{Field3}

  We have investigated the periodic and step-like solutions of the
double-Sine-Gordon equation in this paper. Runge-Kutta algorithm
was employed to integrate the static, second order ODE, with
adjustable initial conditions ($\phi=\pi$ and various
$\frac{d\phi}{dx}$ at $x=0$). The solutions fall into two
categories: periodic, and step-like. Step-like solutions are a
sequence of kinks ($...kkk...$) or anti-kinks
($...\tilde{k}\tilde{k}\tilde{k}...$). Periodic solutions are a
sequence of kink, anti-kink ($...k\tilde{k}k\tilde{k}...$). For
initial conditions considered here, the constant ``pressure'' P
characterizes these solutions with $P>0$ for step-like solutions
and $-2<P<0$ for periodic solutions. We also characterize these
solutions using their energy density and therefore calculate
energy as a function of distance. We are therefore able to
consider an ``interaction energy'' as a function of distance. In
this regard, we find that the step-like solutions of DSG equation
are similar to their SG counterparts, i.e. the behavior of the
system does not change substantially with parameter $\epsilon$.
However, the behavior of periodic solutions depends crucially on
$\epsilon$. In fact, we observe the emergence of an extra branch
of solutions as $\epsilon$ is increased.

Using the concept of ``interaction energy'' we are able to
consider our system as an interacting, many-body system on a one
dimensional chain at zero temperature. Using standard definitions
of pressure and density, we calculate the equation of state for
such a system. In this regard, the step-like solutions show
standard behavior. However, the periodic solutions show unusual
behavior. First, due to the nature of attractive force between
them, they exhibit negative pressure. More interestingly, they
exhibit a transition from a region of positive compressibility to
a region of negative compressibility signifying an intrinsic
instability common in many thermodynamic systems exhibiting phase
transition. This is of particular interest since most one
dimensional interacting systems do not exhibit thermodynamic phase
transition.

Many interesting questions arise: Is there an experimental
realizations for such a system? We believe that given the
prevalence of DSG equation in many experimentally realizable
situations outlined in the introduction, there is hope to study
such systems experimentally and test our results.

Furthermore, other generalization are possible. For example, we
have looked at potentials of the form
$V(\phi)=1+\epsilon-\cos(\phi)-\epsilon \cos(n\phi)$ with $n$
integer larger than $2$, i.e. the multiple Sine-Gordon equation.
Here, instead of two branches we obtain multiple branches. Another
avenue of investigation is the study of inhomogeneity (e.g.
$\epsilon\neq$constant) in these system. Finally, the role of
temperature and fluctuations might be of interest in these
systems. Here, the dynamical stability of the solutions
we have studied here become an important issue.
These issues are currently under investigation and we
intend to report our results in forthcoming publications.

\acknowledgments{N. Riazi  and A. Montakhab acknowledge the
support of Shiraz University. The authors would like to thank A.
Azizi for helpful discussions.}



\begin{thebibliography}{99}
\bibitem{0}M. Ishikawa and K. Hide, J. Phys. C: Solid State
Phys. (1984).
\bibitem{1}R. Khomeriki and J. Leon, Phys. Rev. {\bf E 71}, 056620 (2005).
\bibitem{2} Gaetano Fiore, math-ph/0512002 (2005).
\bibitem{3}N. Riazi, A. Azizi and S. M. Zebarjad, Phys. Rev {\bf D 66},
065003 (2002).
\bibitem{4}L. V. Yakushevich, \textit{Nonlinear Physics of DNA}, Wiley,(2004).
\bibitem{5}L. V. Yakushevich, A. V. Savin and L. I. Manevitch, Phys. Rev. {\bf E 66}, 016614 (2002).
\bibitem{6} Sara Cuenda, Angel Sanchez, Niurka R. Quintero,
Physica {\bf D 223}, 214­221 (2006).
\bibitem{7}J. Timonen, M. Stirland, D. J. Pilling, Yi Cheng, and R. K. Bullough,
 Phys. Rev. Lett. {\bf 56}, 2233 (1986).
\bibitem{8}S. Burdick, M. El-Batanouny and C. R. Willis, Phys.
Rev. {\bf B 34}, 6575 (1986).
\bibitem{9}K. Maki and P. Kumer, Phys. Rev. {\bf B 14}, 118 (1976);
14. 3290 (1976).
\bibitem{10}Y. Shiefman and P. Kumer, Phys. Scr. 20, 435 (1979).
\bibitem{11}K. M. Leung, Phys. Rev. {\bf B 27}, 2877 (1983).
\bibitem{12}O. Hudak, J. Phys. Chem. {\bf 16}, 2641 (1983); {\bf 16}, 2659 (1983).
\bibitem{13}M. El-Batanouny, S. Burdick, K. M. Martini and P.
Stancioff, Phys. Rev. Lett. {\bf 58}, 2762 (1987).
\bibitem{14}E. Magyari, Phys. Rev. {\bf B 29}, 7082 (1984).
\bibitem{15}J. Pouget and G. A. Maugin, Phys. Rev. {\bf B 30}, 5306
(1984); {\bf 31}, 4633(1984).
\bibitem{15b}N. Hatakenaka, H. Takayanagi, Y.Kasai and S. Tanda, Physica {\bf B} 284-288 (2000) 563-564.
\bibitem{16}T. Uchiyama, Phys. Rev. {\bf D 14}, 3520 (1976).
\bibitem{17}S. Duckworth, R. K. Bullough, P. J. Caudrey and J. D.
Gibbon, Phys. Lett. {\bf57 A}, 19 (1976).
\bibitem{18}V. A. Gani and A. E. Kudryavtsev, Phys. Rev. {\bf E 60}, 3305 - 3309
(1999).
\bibitem{19}M. Croitoru, J. Phys. A: Math. Gen. {\bf22}, 845-863 (1989).
\bibitem{20} Constantine A. Popov, Wave Motion. {\bf42(1)}, 309-350 (2006).
\bibitem{21}N. Riazi and A. R. Gharaati, Int. J. Theor. Phys. {\bf37},
1081 (1998).
\bibitem{21b} M. Wang and X. Li, Chaos, Solitons and Fractals.
{\bf27}, 477 (2006).
\bibitem{21c} C. A. Condat, R. A. Guyer and M. D. Miller, Phys. Rev. {\bf B 27
}, No. 1, 474 (1983).
\bibitem{22}L. H. Ryder, \textit{Quantum Field Theory}, Cambridge University Press (1985).
\bibitem{23}M. Guidry, \textit{Gauge Field Theories, an Introduction with
Applications}, Wiley, NewYork (1991).
\bibitem{24}N. Riazi, Int. J. Theor. Phys. GTNO, {\bf 8}, No. 2, 115 (2001).
\bibitem{24a}E. Kreyszig, \textit{Advanced Engineering Mathematics}, John
Wiley and Sons, NewYork(1983).
\bibitem{24b}The lower bound for the value of $P$, in general,
depends on $\epsilon$. However, for the set of initial conditions
we consider this lower bound is $\epsilon$ independent and is
equal to $-2$.
\bibitem{25a}Note that from Eq.(\ref{qq}), $\rho_{v}=T^{0}_{0}=2=$constant for the false vacuum and
$E\approx\rho_{v}\times L$ at large $L$.
\bibitem{25}R. D'Inverno, \textit{Introducing Einstein's  Relativity},
Oxford University Press, NewYork (1992).
\bibitem{26}H. B. Callen, \textit{Thermodynamics and an Introduction to Thermostatistics},
John Wiley and Sons, NewYork (1985).
\end{thebibliography}
\end{document}